\documentclass[aps,prl,reprint,superscriptaddress,twocolumn]{revtex4-1} 

\usepackage{graphicx}
\usepackage{longtable}
\usepackage{color}
\def\lsim{\, \rlap{$<$}{\lower 1.1ex\hbox{$\sim$}}\,}
\begin{document}

%\preprint{YNUCOS1710}

\title{Primordial Perturbation of Dark Matter as a Novel Probe of Very Early Universe}

\author{Changhong~Li}
\email{changhongli@ynu.edu.cn}
\affiliation{Department of Astronomy,  Key Laboratory of Astroparticle Physics of Yunnan Province,\\ Yunnan University, No.2 Cuihu North Road, Kunming, China 650091}

\begin{abstract}
Dark matter(DM) is the only possible candidate which would be apart from the thermal equilibrium before Big Bang nucleosynthesis(BBN) in accordance with current DM searches. In this work, we report a generic scenario that primordial perturbation of dark matter(PPDM) can be, effectively, generated and encoded with primordial information of very early universe up to the reheating era. We present an analytical solution of the whole evolution of PPDM. A novel and strong constraint on the reheating process imposed by primordial gravitational wave(PGW) is obtained for the first time. It indicates the ratio of PGW to primordial curvature perturbation(PCP) is not only dependent on the slow-roll spectral index but also, strongly, on the decay process of inflaton at reheating. For very generic reheating process, our result provides a natural explanation of the paucity of PGW in current observations.
\end{abstract}

\pacs{170901}

\maketitle
\section{Introduction}
One of the most interesting intellectual challenges in recent is how to unveil the early history of universe by utilizing an array of observable probes such as cosmic microwave background(CMB) anisotropies and dark matter(DM).  

The precisely measured CMB anisotropies\cite{Komatsu:2010fb,Ade:2015xua} has almost crowned inflation\cite{Guth:1980zm} except the primordial gravitational wave(PGW) has not been detected\cite{Ade:2014xna} as optimistic estimation\cite{Dodelson:2003ft}. Another example of the current observations not warranting the optimistic estimation is that DM has not been detected by the highly precise direct DM searches\cite{Fu:2016ega}. 

In this work, we conjecture that these two issues are related, in presence of DM production, by the reheating process\cite{Bassett:2005xm}. The insight motivates us is that the variation of primordial curvature perturbation(PCP) during the particle productions is roughly inverse to the total thermally averaged cross section(TTACS) of particles,
\begin{equation}
\Delta \Phi_\alpha\sim   \langle\sigma v\rangle_\alpha^{-1}~, 
\end{equation}
where $\alpha$ labels different species of particles. In general, TTACS of all known Standard Model(SM) particles are relatively large as they are already detected at LHC~\cite{Aad:2012tfa, Chatrchyan:2012xdj}. So, for SM particles, this effect is almost indiscernible in the measurement of CMB anisotropies~\cite{Martin:2014nya, Martin:2010kz}. It makes the constraints on reheating imposed by CMB is relatively weak.

The novel exception is DM. In accordance to recent (in-)direct DM searches including at LHC\cite{Aprile:2013doa,Khachatryan:2014rra,Akerib:2016lao,Tan:2016zwf,Gordon:2013vta}, $\langle\sigma v\rangle_\chi\sim 0$ in comparing with all SM particles. It may lead some observable signals on CMB. 

In this work, the analytical solution of $\Delta\Phi_\chi$ is obtained by investigating the generation of primordial perturbation of dark matter(PPDM). A novel and strong constraint on the reheating process imposed by $\Phi$ (virtually, by PGW) is, accordingly, obtained for the first time.  It turns out the value of $\Delta\Phi_\chi$ is strongly dependent on the initial abundance of DM at the end of reheating process. For very generic reheating process, our result indicates  
\begin{equation}
\Delta\Phi_\chi  \gg \Phi_\varphi~,
\end{equation}
where $\Phi_\varphi$ is PCP generated during inflation. Therefore, it provides a very natural explanation of the smallness of PGW in consistent with current observations.

To sum up, this mechanism we report in this work is not only valid for DM produced in thermal equilibrium~\cite{Kolb:1990vq,Feng:2008ya} and also for in non-thermal~\cite{Chung:1998ua, Li:2014era, Feldstein:2013uha}. We perform our analysis in both scenarios. We find $\Delta \Phi_\chi$ is much smaller in thermal equilibrium scenario than in non-thermal-equilibrium scenario. Since $\langle\sigma v\rangle_\chi$ is, generically, much larger in thermal equilibrium than in non-thermal-equilibrium scenarios, this result confirms our original expectation.

\section{Primordial Perturbation of Dark Matter in Early Universe}
We focus on the scenario that DM candidate, $\chi$, is produced thermally in early universe with the interaction, $ \varphi\rightarrow\phi+\phi\leftrightarrow \chi+\chi $~, where $\varphi$ is the quantum field  driving inflation earlier and $\phi$ would be one of SM particles or mediators~\cite{Dodelson:2003ft, Mukhanov:2005sc, Peskin:1995ev}. To facilitate the analysis of cosmic evolution of long wavelength PPDM, we start with the unintegrated Boltzmann equation, $\frac{df}{dt}=C[f]$, where $C[f]$ is the collision term,  $f$ is the distribution function $f=e^{\frac{\mu}{T}}e^{-\frac{E(p)}{T}}\left[1-\Theta(\vec{x}, t)\right]$ and $\Theta(\vec{x}, t)$ is its perturbation induced by the metric perturbation. In the perturbed FRW metric, $g_{\mu\nu}=\{-1-2\Psi(\vec{x},t), a^2(t)\delta_{ij}\left[1+2\Phi(\vec{x},t)\right]\}$,  we decompose the unintegrated Boltzmann equation into the {\it zeroth order}~\cite{Dodelson:2003ft},
\begin{equation}\label{eq:zoe}
\frac{d n_\chi}{dt}+3Hn_\chi=\widetilde{\langle \sigma v\rangle}\left(n_\phi^2-n_\chi^2\right)~, 
\end{equation}
and the {\it first order}, 
\begin{equation} \label{eq:dtheta}
\frac{d \Theta}{dy}-3\frac{d\Phi}{dy}=\frac{\widetilde{\langle \sigma v \rangle}}{ H y n_\chi}\left(n_\phi^2-n_\chi^2\right)\left(\Theta+\Phi\right)~,
\end{equation}
where $\widetilde{\langle\sigma v\rangle}$ and $m_\chi$ are, respectively, the thermally averaged cross section and mass of DM,  $n_\chi$ and $n_\phi$ the number density of $\chi$ and $\phi$, $\mu$ and $T$ the chemical potential and temperature, $B$ is short for $B_k$ for the long wavelength modes of each perturbative quantity,  the terms involving $\frac{\partial ~}{\partial x^i}$ are discarded and $\Phi=-\Psi$ is taken in the long wavelength limit, and  $y\equiv \frac{m_\chi}{T}$ is adopted in the thermal bath of early universe after reheating. Accordingly, the long wavelength modes of PPDM take
 $\delta\rho_\chi=-\rho_\chi\Theta$. And $\Phi$ and $\Theta$ are also the governed by perturbed Einstein equation, $\delta G^0_0=8\pi G\delta T^0_0$, which leads
 \begin{equation}\label{eq:efo}
\frac{1}{H}\frac{d\Phi}{dt}+\Phi=-\frac{1}{2}\left[\left(1-\gamma\right)\Theta-\gamma\frac{\delta\rho_\varphi}{\rho_\varphi}\right]~,
\end{equation}
where $\rho_\varphi$ and $\delta\rho_\varphi$ are the energy density and its perturbation of $\varphi$, $\rho_r$ is the total energy density of all possible relativistic components, and $\gamma$ is ratio of $\rho_\varphi$ to the total energy density of universe, $\gamma\equiv\frac{\rho_\varphi}{\rho_r+\rho_\varphi}$\footnote{Inflaton $\varphi$ may decay into other SM particles. In this work, we focus on the assumption that DM is only produced by $\phi$. So DM is, essentially, irrelevant to other SM particles except their productions may affect the evolution of background temperature before the end of reheating. And such effect is packaged in the key parameter of reheating -- the reduced initial abundance of DM at the end of reheating -- $\xi$.}.  

\begin{figure}[htp!]
\centering
\includegraphics[width=0.48\textwidth]{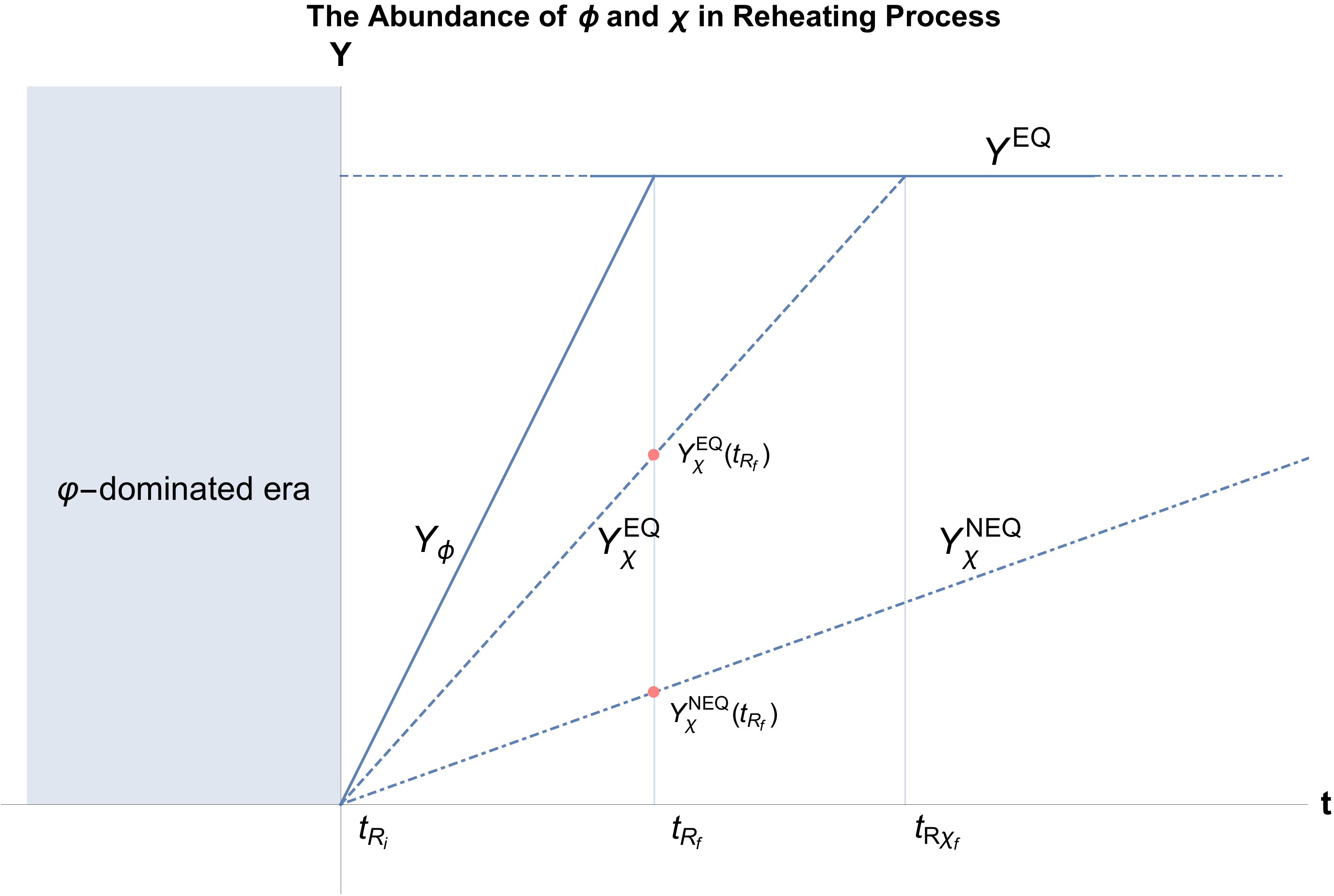}
\caption{A schematic plot of the evolution of abundance of $\phi$ and $\chi$ during reheating. $t_{R_i}$ is the onset of $\phi$ production. $t_{R_f}$ and $t_{R\chi_f}$ are the moments attaining thermal equilibrium for $\phi$ and $\chi$ respectively. $Y_\chi(t_{R_f})$ is the abundance of DM at the end of the reheating process, and the superscripts, $~^{EQ}$ and $~^{NEQ}$, label the thermal equilibrium and non-thermal-equilibrium scenarios respectively. }
\label{fig:reheating.pdf}
\end{figure}

In general, universe is reheated very swiftly\cite{Kofman:1994rk} so that $\Phi$ and $\frac{d\Phi}{dt}$ are fixed during reheating according to Eq.(\ref{eq:efo}),
\begin{equation}~\label{eq:mcr}
\Delta\left(\Phi\right)_R\rightarrow 0~,\quad \Delta\left(d\Phi/dt\right)_R\rightarrow 0~,\quad \text{if}\quad \Delta t_R\rightarrow 0~,
\end{equation} 
where the subscript $_R$ labels the whole of reheating process between $t_{R_i}$ and $t_{R_f}$, and $_{R_i}$ and $_{R_f}$ label the onset and end of $\phi$ production through $\varphi\rightarrow \phi$. The reheating process is plotted in FIG.\ref{fig:reheating.pdf} schematically. $t_{R\chi_f}$ is the moment of DM attaining thermal equilibrium in thermal equilibrium scenario, and in non-thermal-equilibrium scenario, $t_{R\chi_f}\rightarrow \infty~$, accordingly. 

\subsection{In Thermal Equilibrium Scenario}
In thermal equilibrium scenario, after reheating, $\varphi$ is drained, $\gamma=0$, and the thermal equilibrium is attained, $n_\chi=n_\chi^{EQ}$.  According to Eq.(\ref{eq:dtheta}), Eq.(\ref{eq:efo}) and Eq.(\ref{eq:mcr}), $\Theta(y)$ induced by $\Phi(y)$ is fully determined by $\Phi_\varphi$, which leads the well-known relation\cite{Dodelson:2003ft},
\begin{equation}\label{eq:tpeq}
\Theta(y)=-2\Phi_\varphi~,
\end{equation}     
where $\Phi_\varphi$ is PCP generated during inflation and $d\Phi_\varphi(t_{R_i})/dt =0$ is taken for inflation. As $n_\chi=n_\chi^{EQ}$ in thermal equilibrium, the PPDM, $\delta\rho_\chi=6 T^4\pi^{-2} \Phi_\varphi$, carries no information about reheating, just like other relativistic components\footnote{For simplicity, we take the degeneracy of the species of each particle of interest equal to 1, $g_\chi=g_\phi=1$, in this work. It is not difficult to derive similar result for different values of $g_\chi$ and/or $g_\phi$.}.     

We may have noticed that, to derive Eq.(\ref{eq:tpeq}),  DM is also assumed to attain thermal equilibrium very swiftly, $(t_{R\chi_f}-t_{R_f}) \rightarrow 0$, which is valid for a number of models in this scenario. However, DM can take a relatively long time to attain thermal equilibrium for other models in this scenario\cite{Li:2014era}. Therefore, in general, we have
\begin{equation}\label{eq:thieq}
\Theta(y)=-2\Phi_\varphi-2\Delta\Phi~,
\end{equation}
where the value of $\Delta\Phi$ can be estimated, 
\begin{equation}~\label{eq:vauc}
\Delta\Phi=- y_{R\chi_f}^{-1} \left[\frac{1}{2} \int_{y_{R_f}}^{y_{R\chi_f}}\Theta dy+(y_{R\chi_f}-y_{R_f})\Phi_\varphi\right],
\end{equation}
by utilizing Eq.(\ref{eq:efo}). It leads serval comments: 1) $\Delta\Phi$ is a constant since thermal equilibrium is attained eventually in this scenario, and $\Delta\Phi\rightarrow 0$ if $(y_{R\chi_f}-y_{R_f}) \rightarrow 0$; 2) $\Delta\Phi$ is encoded with the information of early universe from $y_{R_f}$ to $y_{R\chi_f}$ since it is affected by the generation of PPDM during this period; 3) $\Delta\Phi$ also carries the information from $t_{R_i}$ to $t_{R_f}$, {\it i.e.} the reheating era, as it is strongly dependent on the value of $\Theta(y_{R_f})$. This point will become manifest in detailed analysis in Appendix.

\subsection{In Non-thermal-equilibrium Scenario}

In non-thermal-equilibrium production scenario, after reheating, $\varphi$ is also drained, $\gamma=0$, and the abundance of DM is much lower than the thermal equilibrium, $n_\chi\ll n_\chi^{EQ}$. Since the abundance of DM is not enveloped by thermal equilibrium, the exact evolution of its abundance and perturbation must be sensitive to cosmic evolution and nature of DM candidates, {\it i.e.} model-dependent~\cite{Li:2014cba}.  

In this work, we consider the simplest case that DM is produced thermally with the minimal coupling, $\mathcal{L}_{int}=\lambda \phi^2\chi^2$~, for illustrative purpose. After reheating, the abundance of DM  can be obtained by solving Eq.(\ref{eq:zoe}) with the initial condition, $Y(y_{R_f})\equiv\kappa\xi$, 
\begin{equation}\label{eq:yin}
Y\equiv n_\chi T^{-3}=\kappa(y-y_{R_f}+\xi)~,
\end{equation}
where $\kappa\equiv m_\chi^3\langle \sigma v\rangle (4\pi^4H y^2)^{-1}$, $\langle\sigma v\rangle$ and $\bar{\langle\sigma v\rangle}$ are the value of $\widetilde{\langle\sigma v\rangle}$ in low and high temperature respectively, $\bar{\langle\sigma v\rangle}=\langle\sigma v\rangle\frac{y^2}{4}$ in the limit $m_\phi\ll m_\chi$~\cite{Peskin:1995ev}, $\mu_\phi=0$ after reheating, and $\kappa\ll (2\pi^2)^{-1}$ in this scenario~\cite{Li:2014era}.

Remarkably, with Eq.(\ref{eq:zoe}), the reduced initial abundance of DM at the end of reheating $t=t_{R_f}$~, 
\begin{equation}\label{eq:xiform}
\xi= \kappa^{-1}T_{R_f}^3\int_{t_{R_i}}^{t_{R_f}}\widetilde{\langle\sigma v\rangle} n_\phi^2 dt~,
\end{equation}
becomes a key parameter of reheating, which plays a central role in whole scenario.  It consists of  three very important issues in early universe: 1) the nature of DM, $\langle\sigma v\rangle$ and $m_\chi$; 2) the typical temperature, $T_{R_f}$, and duration, $t_{R_i}\rightarrow t_{R_f}$, of reheating ; and 3) how does the inflaton decay, $\varphi\rightarrow \phi+\cdot\cdot\cdot$, and how does the decay rate be corrected with finite temperature, {\it i.e.} the explicit form of $n_\phi$ and $\widetilde{\langle\sigma v\rangle}$ in fast varying background temperature. Therefore, once $\xi$ is constrained by the observations such as the detection of PGW discussed in next section, the details of inflaton decay and nature of DM particles can be essentially unveiled. And such result can be crosschecked by the (in)-direct DM searches and/or at LHC, hopefully, in near future.

Generically, the reheating takes place very swiftly so that the value of $\xi$ should be very small, $\xi\ll 1$. And $y_{R_f}$ is almost negligible as the reheating temperature is very high in generic\cite{Bassett:2005xm}. Therefore, we can focus on the parameter region, $y_{R_f}\ll \xi \ll 1$, through the following analysis\footnote{The full parameter region viable in non-thermal-equilibrium scenario takes $y_{R_f}\ll \xi \ll (2\pi^2\kappa)^{-1}$. However, $1\ll\xi \ll (2\pi^2\kappa)^{-1}$ implies a unusual long time for the reheating process, which leads a trivial result, $\Phi(y)=\Phi_\varphi$, after freezing-out.}.   

By solving Eq.(\ref{eq:dtheta}) with Eq.(\ref{eq:efo}) and Eq.(\ref{eq:yin}) and matching the solution at $y_{R_f}$ with Eq.(\ref{eq:mcr}), we obtain
\begin{equation}\label{eq:phineq}
\Phi(y)=\Phi_\varphi\times \mathcal{G}\left(-y\times\xi^{-1}\right)~,
\end{equation}
where $\mathcal{G}(x)\equiv1+\sum_{k=1}^{\infty}\frac{(\frac{3-\sqrt{17}}{4})_k(\frac{3+\sqrt{17}}{4})_k}{(\frac{7}{2})_k}\frac{x^k}{k!}$ and $(u)_k\equiv\Pi_{n=0}^{k-1}(u+n)$.

\paragraph{\bf After freezing-out}
After freezing-out, $T\ll m_\chi$, the abundance of DM is frozen, $Y_{fo}=Y(y=1)=\kappa$.  
By imposing the current value of $\Omega_\chi$, $\Omega_\chi=\mathcal{A} m_\chi Y_{fo}=0.26$ in which $\mathcal{A}$ is a constant parameter, the cosmological constraint on $m_\chi$ and $\langle\sigma v\rangle$ is obtained~\cite{Cheung:2014nxi}, $\langle\sigma v\rangle=14.4\times 10^{-26}m_\chi^{-2}$. The analysis of detectability in~\cite{Cheung:2014pea} indicates that such DM candidate has not been excluded by current DM searches~\cite{YKECheung:2016zra}.   

In this work, we focus on the evolution of $\Phi$ and $\Theta$. The long wavelength modes of interest will re-enter horizon after DM freezes out. After freezing-out, the collision term vanishes which leads $\Phi(y)=C_1y^{-\frac{5}{2}}+C_2$. By matching it with Eq.(\ref{eq:phineq}) at $y=1$, we obtain,
\begin{equation}\label{eq:phifo}
\Phi=\Phi_\varphi\mathcal{G}\left(-\xi^{-1}\right)\left[1-\frac{\Gamma\left(\frac{3+\sqrt{17}}{4}\right)}{5 \Gamma\left(\frac{7+\sqrt{17}}{4}\right)}\right]
\end{equation} 
from freezing-out to matter-radiation equality, where $\xi\ll 1$ and  $y\gg 1$ are taken and $\Gamma(x) \equiv \int_0^\infty z^{x-1}e^{-z}dz$. Encouragingly, $\Phi$ has a same wave vector dependence with $\Phi_\varphi$, which is favored by the current measurement of the slow-roll spectral index\cite{Komatsu:2010fb}.     

Accordingly, a novel and falsifiable result of PPDM is obtained,  
\begin{equation}\label{eq:drf}
\delta\rho_\chi=\frac{\langle\sigma v\rangle m_\chi^2 }{2\pi^4 }\times\frac{T^5}{H}\times \Phi~,
\end{equation} 
where $\delta\rho_\chi$ is related to initial distribution of DM\footnote{Hopefully, by taking account its evolution after matter-radiation equality, PPDM ($\delta\rho_\chi\sim k^{-5}$) obtained in this work may also shed some new insights on formation of large-scale structure and primordial black hole\cite{Li2017pre}.}, $\langle\sigma v\rangle$ and $m_\chi$ can be, hopefully, determined by (in-)direct DM searches in near future, and $\Phi$ has already be, essentially, measured by current observations of CMB. 

To sum up, since the analysis of $\Phi$ going through matter-radiation equality and afterward is very mature, $\Phi$ can be traced back from current observations to the moment ahead of matter-radiation equality (see~\cite{Dodelson:2003ft, Mukhanov:2005sc,Liddle:2000cg} for nice textbooks). Therefore, we complete our analysis with Eq.(\ref{eq:phifo}) and, accordingly, Eq.(\ref{eq:drf}) as they are already able to be justified/falsified by astrophysical observations such as the searches of PGW and DM.

\section{A  Constraint on reheating imposed by Primordial gravitational wave}
\begin{figure}[htp!]
\centering
\includegraphics[width=0.48\textwidth]{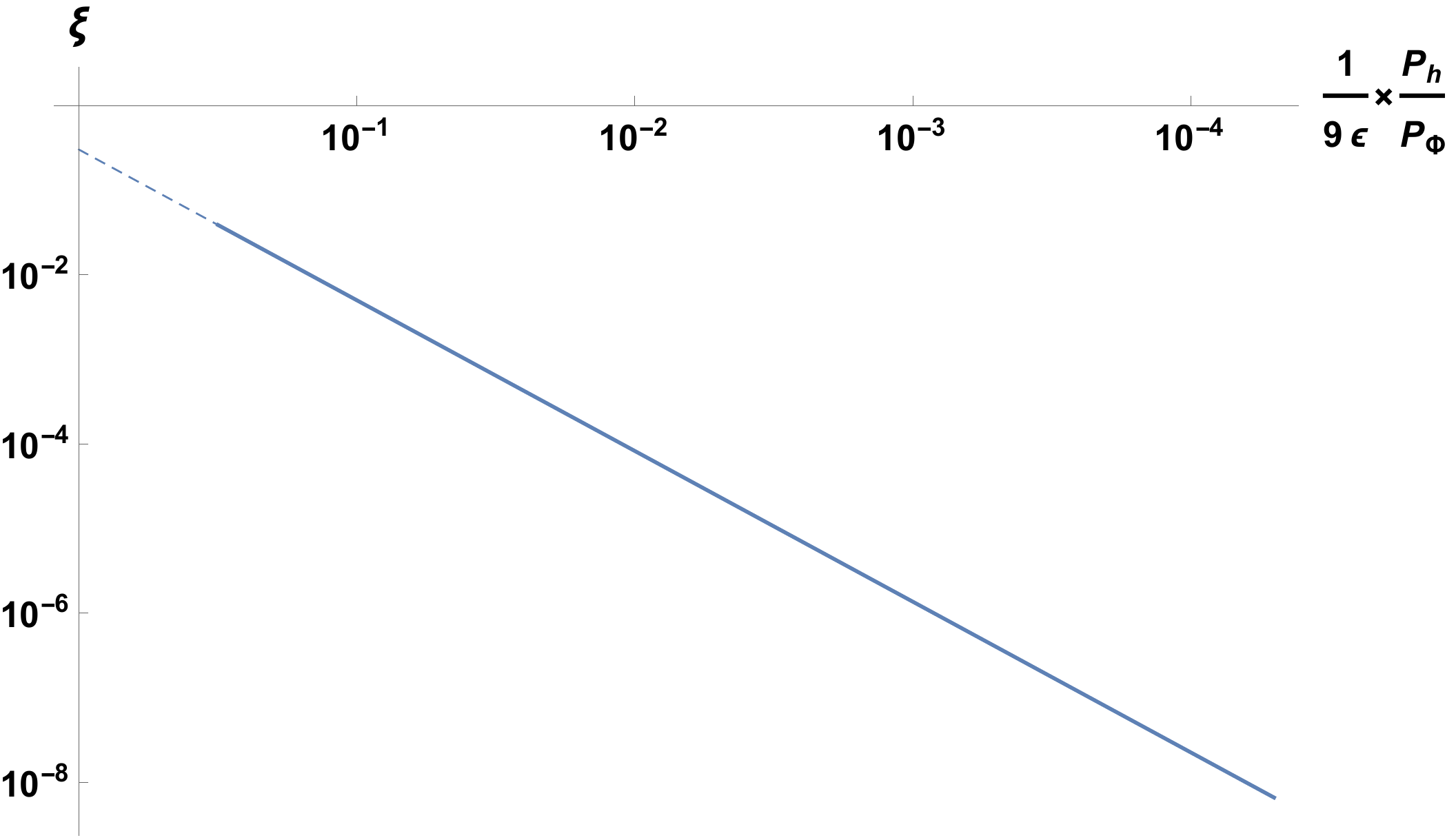}
\caption{The constraints on the key parameter of reheating, $\xi$, imposed by the searches of PGW, $P_h$. Given the value of $\epsilon$ and $P_\Phi$ measured by WMAP and Planck\cite{Komatsu:2010fb,Ade:2015xua}, roughly, $\xi>0.1$ have been excluded by BICEP2\cite{Ade:2014xna}.}
\label{fig:xiprg.pdf}
\end{figure}

In standard cosmology, the spectra of PGW takes
\begin{equation}
P_h=9\epsilon P_{\Phi_\varphi}
\end{equation}
where $P_{\Phi_\varphi}$ is the spectra of scalar mode of PCP generated by quantum fluctuation of $\varphi$ during inflation, $P_B\propto \langle B\rangle^2$, and $\epsilon$ is the slow-roll parameter, $\epsilon\equiv \frac{d}{dt}\left(\frac{1}{H}\right)$~\cite{Dodelson:2003ft}. With Eq.(\ref{eq:phifo}), a novel and strong constraint on the reheating imposed PGW is obtained analytically,
\begin{equation}\label{eq:phreh}
P_h=9\epsilon \left[0.888\times\mathcal{G}\left(-\xi^{-1}\right)\right]^{-2}P_\Phi~,
\end{equation} 
where $1-\Gamma\left(\frac{3+\sqrt{17}}{4}\right)/5 \Gamma\left(\frac{7+\sqrt{17}}{4}\right)=0.888$ is taken. It indicates this key parameter of reheating, $\xi$, can be precisely determined in PGW searches as shown in FIG.\ref{fig:xiprg.pdf}, since $P_\Phi$ and $\epsilon$ have been measured by WMAP and Planck~\cite{Komatsu:2010fb,Ade:2015xua}. Eventually, it can constrain various inflation models with decay rate of inflaton -- in additional to the traditional approach that inflation is testified during slow-roll epoch in measuring slow-roll spectral index. 
 
Since $\mathcal{G}\left(-\xi^{-1}\right)\gg1$ for a generic reheating process in which $\xi\ll 1$, we have
\begin{equation}
P_h\ll 9\epsilon P_{\Phi}~,
\end{equation} 
which provides a natural explanation of the paucity of PGW in recent searches\cite{Ade:2014xna}.

\section{Summary}
In this work, we report, after reheating, the long wavelength modes of PCP are enhanced by DM production. Such enhancement is strongly dependent on the initial abundance of DM at the end of reheating, which strongly constrains the reheating process. Our prediction of $\Phi$, which has the same wave vector dependence with $\Phi_\varphi$, is favored by current observation of CMB anisotropies\cite{Komatsu:2010fb,Ade:2015xua}. 

A novel and strong constraint on reheating process imposed by PCP is obtained in Eq.(\ref{eq:phifo}) (virtually by PGW in Eq.(\ref{eq:phreh})) for the first time by investigating the whole evolution of PPDM analytically.  Encouragingly, it provides a compelling and promising approach to constrain the decay process of inflaton, which is additional to the traditional approach that inflation is testified during slow-roll epoch in measuring spectral index.   

Furthermore, for very generic reheating process in which $\xi\ll 1$, PGW is much smaller than naive expectation, $P_h\ll 9\epsilon P_\Phi$~, which provides a very natural explanation of the paucity of PGW in recent searches.

In prospect, once PGW is detected in very near future, combing with $P_\Phi$ and $\epsilon$ which have been measured by WMAP and Planck, the key parameter of reheating, $\xi$, will be precisely determined, which will constrain the decay process of inflaton strongly. And such result can be crosschecked by the (in-)direct DM searches and at LHC\cite{Aprile:2013doa,Khachatryan:2014rra,Akerib:2016lao,Tan:2016zwf,Gordon:2013vta}.

\section{acknowledgments}
The work has been supported in parts by the National Natural Science Foundation of China (11603018), the FRYPMST Grants(2016FD006), the National Natural Science Foundation of China (11433004,11690030), the Leading Talents of Yunnan Province (2015HA022) and the Top Talents of Yunnan Province (2015HA030).

\section{Appendix}
In thermal equilibrium scenario, $\Delta \Phi$ in Eq.(\ref{eq:thieq}) is mainly generated during the period of $n_\chi\ll n_\chi^{EQ}$, which can be approximated by utilizing Eq.(\ref{eq:phineq})\footnote{We also, accordingly, take $\mathcal{L}_{int}=\lambda \phi^2\chi^2$ for illustration. In thermal equilibrium scenario, $2\pi^2\kappa \gg 1$ and the abundance of DM is $Y=\frac{(1-e^{-2\pi^2\kappa(y-y_{R_f}+\xi)})}{[\pi^2(1+e^{-2\pi^2\kappa(y-y_{R_f}+\xi)})]}$. Approximately, $Y\simeq\kappa(y-y_{R_f}+\xi)$ during $y_{R_f}\le y<y_{R\chi_f}$. },
$\Delta \Phi\simeq\Phi_\varphi\left[\mathcal{G}(-y_{R\chi_f}\xi^{-1})-1\right]$. With $2\pi^2\kappa y_{R\chi_f}\simeq 1$, we have 
\begin{equation}
\Delta \Phi\simeq\Phi_\varphi\left[\mathcal{G}\left(-(2\pi^2 \kappa \times\xi)^{-1}\right)-1\right]~.
\end{equation}
Since thermal equilibrium is attained before freezing-out, we have $2\pi^2\kappa\gg 1$ in this scenario. Therefore, for a given value of $\xi$, the variation of $\Phi$ after $y_{R_f}$ is much smaller in thermal equilibrium scenario than in non-thermal-equilibrium scenario, which confirms our original expectation. 

\begin{figure}[htp!]
\centering
\includegraphics[width=0.48\textwidth]{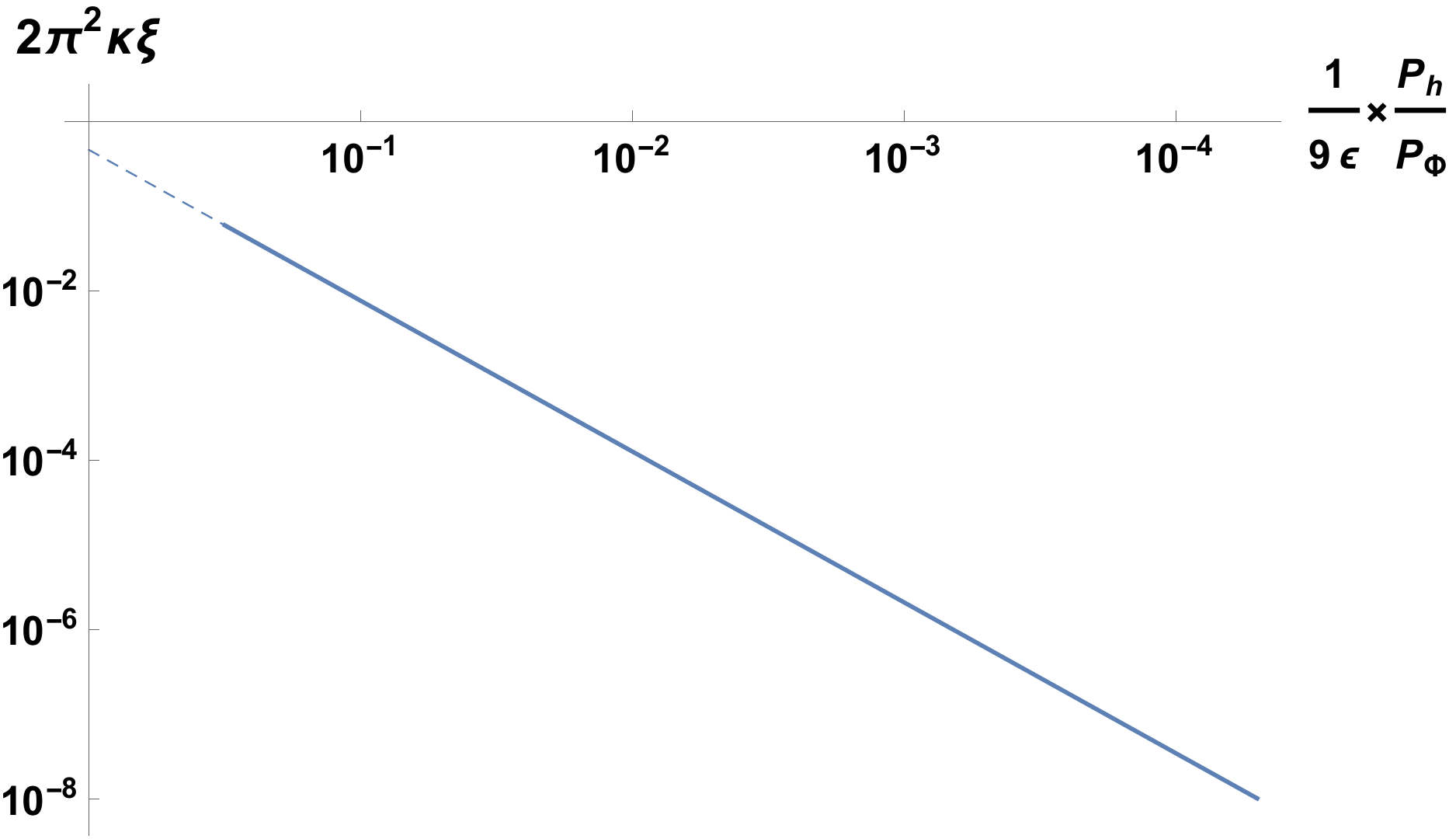}
\caption{The constraints on $2\pi^2\kappa \xi$ imposed by $P_h$ in thermal equilibrium scenario. }
\label{fig:Xiprgeq.pdf}
\end{figure}

Accordingly, PPDM takes 
\begin{equation}
\delta\rho_\chi\simeq6 T^4\pi^{-2} \Phi_\varphi\mathcal{G}\left(-(2\pi^2 \kappa \times\xi)^{-1}\right)~,
\end{equation}
and the constraint on the reheating with PGW becomes
\begin{equation}
P_h\simeq9\epsilon \left[\mathcal{G}\left(-(2\pi^2\kappa \xi)^{-1}\right)\right]^{-2}P_\Phi~
\end{equation}
in thermal equilibrium scenario as shown FIG.\ref{fig:Xiprgeq.pdf}. In the parameter region, $2\pi^2\kappa \xi\ll 1$, we also have $P_h\ll 9\epsilon P_\Phi$ in the thermal equilibrium scenario. 

% Create the reference section using BibTeX:
%\bibliography{BBGref}
%\input{BBG.bbl}

\end{document}